\documentclass[onecolumn]{aastex631}

\usepackage{xcolor}
\usepackage{subfloat}
\usepackage{multirow}
\usepackage{soul}
\usepackage{appendix}
\usepackage{graphicx}	
\usepackage{amsmath}	

\usepackage{amssymb}	
\usepackage{cleveref}   
\usepackage{ulem}
\usepackage{multirow}
\usepackage{placeins}

\shorttitle{Kepler-396 Planetary System}
\shortauthors{Yeh, Jiang et al.}

\begin{document}

\title{On the Existence of a Third Planet and an  
Apsidal Resonance Chain in the Kepler-396\\ System through TESS Data}

\author[0000-0001-8677-0521]{Li-Chin Yeh}
\affiliation{Institute of Computational and Modeling Science, National Tsing Hua University, Hsinchu 30013, Taiwan}
\email{lichinyeh@mx.nthu.edu.tw}
\author[0000-0001-7359-3300]{Ing-Guey Jiang}
\affiliation{Institute of Astronomy, National Tsing Hua University, Hsinchu 30013, Taiwan}
\affiliation{Department of Physics, National Tsing Hua University, Hsinchu 30013, Taiwan}
\email{jiang@phys.nthu.edu.tw}
\author[0009-0000-0807-2170]{Prangsutip Cherdwongsung}
\affiliation{Institute of Astronomy, National Tsing Hua University, Hsinchu 30013, Taiwan}
\author[0000-0002-8988-8434]{D. Bisht}
\affiliation{Indian Centre For Space Physics, 
466, Barakhola, Singabari road, Netai Nagar, Kolkata, West Bengal, 700099, India}
\author[0009-0003-8446-4557]{Shraddha Biswas}
\affiliation{Indian Centre For Space Physics, 
466, Barakhola, Singabari road, Netai Nagar, Kolkata, West Bengal, 700099, India}
\author[0009-0003-9593-7460]{Kaviya Parthasarathy}
\affiliation{Institute of Astronomy, National Tsing Hua University, Hsinchu 30013, Taiwan}
\author{Hsin-Min Liu}
\affiliation{Institute of Astronomy, National Tsing Hua University, Hsinchu 30013, Taiwan}

\begin{abstract}
The possible existence of a third planet in the Kepler-396 system has been previously proposed. 
Given that the planetary mass is poorly constrained 
and the orbit could be either interior or exterior to planet Kepler-396 c, that proposal is not clearly known 
in the community.
In order to address this and further investigate Kepler-396 system's configuration, new TESS data are employed to form a long baseline of transit timing. 
Using the pixel-level decorrelation technique, the mid-transit times are determined.
Through long baseline transit-timing-variation
technique, 
we perform a comprehensive exploration of the parameter space of the third planet. 
Two best-fit solutions, derived from different priors
of orbital inclinations, have led to the reduced 
chi-square values as 2.24 and 2.34 respectively,
compared to 12.97 for the two-planet model.
This result strongly suggests the
existence of the third planet, i.e. Kepler-396 d.
Based on these two best-fit solutions, 
the masses of Kepler-396 b, c, d are updated to be around
9, 4, 4 $M_\earth$,
and the corresponding orbital periods are
43, 89, 144 days, respectively.
With these updated masses, 
Kepler-396 b is on the border
between rocky or volatile-rich, 
Kepler-396 c is confirmed to be a volatile-rich super-puff planet, and both planets 
lie in the Neptune savanna. 
Dynamical simulations with three planets are performed, showing that
they are stable for a timescale of $10^6$ years. The examination on
resonant angles leads to the conclusion that planets Kepler-396 b, c, d form a three-body apsidal resonance chain.
\end{abstract}

\keywords{planetary systems --- planet and satellites: individual (Kepler-396 b) ---  planet and satellites: individual (Kepler-396 c) --- techniques: photometric}

\section{Introduction}
\label{sec:intro}

During the past three decades, remarkable
growth in the field of exoplanet science has been triggered by the continuous flow of new detections, which started from the discoveries of hot Jupiters through the radial velocity method \citep{MayorQueloz1995}. 
These results were strengthened by later confirmation and further exoplanet detections through transit photometry observations.
These transit surveys were first performed through ground-based telescopes \citep{Alonso+2004ApJ, Bakos+2004PASP, Pollacco+2006PASP}, which were followed by the later effort of transit observations such as {\it The Transit Light Curve Project} \citep{Holman2006ApJ, Winn2007AJ}.
Transit space missions, including CoRoT \citep{Auvergne2009}, Kepler \citep{Borucki2010Sci}, TESS \citep{Ricker2015TESS}, and CHEOPS \citep{Benz2021}, continue to fuel further
rapid development.
With high-precision photometry, Kepler detected many more
sub-Neptunes and super-Earths and found that these are dominant populations among all discovered transit exoplanets \citep{Latham2011, Dorn2015}. In addition, the Kepler space mission also leads to the discovery of many more multi-planet systems \citep{Ciardi2013, Millholland2017}. 
Among these systems, there is an excess of planet pairs associated with mean-motion resonances \citep{Lissauer2011}. Thus, \cite{Delisle2014} started to investigate possible mechanisms causing the formation of pair planets near resonance.  Besides, \cite{Barnes2015} later studied chaotic orbits in these systems. 
The statistic result presented in \cite{Leleu2024} indicated that those exoplanets in/near mean motion resonances have smaller densities than those far from mean motion resonances, indicating two different formation histories. In addition, some planets have been found to be
in/near a resonant chain in these multi-planet systems \citep{Christiansen2018}.
Finally, the occurrence rate of super Earths was found to be higher when there are cold or warm Jupiters in the
system \citep{ZhuWu2018, Gajendran2024}. 

Inspired by the above exciting and fruitful discoveries, many more follow-up transit observations on these known
exoplanets were performed in order to refine and update important orbital elements.
These later efforts gradually build up a longer baseline
of transit timing over epochs, which leads
to the detection of possible transit timing variations (TTVs)
\citep{Jiang2013AJ, Jiang2016AJ, Gajdo2017, 
Southworth2019}.
In particular, \cite{Blecic+2014ApJ} and \cite{Jiang2016AJ} proposed that the
potential TTVs of WASP-43 b, a hot Jupiter detected by \cite{Hellier2011}, could be due to an orbital decay. 
Their results also provided an updated stellar dissipation coefficient, which lies within the theoretically allowed range. 
Though the orbital decay of WASP-43 b is under debate and 
has never been confirmed \citep{Hoyer+2016AJ, 
Garai+2021MNRAS,
Davoudi+2021AJ}, this proposal triggered 
further investigations on many other systems 
and led to a confirmed case
such as WASP-12 b 
\citep{Maciejewski+2016AA, Patra+2017AJ, Yee+2020ApJ}. 
In addition, in order to explain TTVs in various systems \citep{Bouma+2019AJ, Mannaday+2020, Sariya+2021RAA, Su+2021AJ, Maciejewski+2021A&A, Mannaday+2022,
A-thano+2022AJ, A-thano+2023, Yang+2024MNRAS, Biswas+2024, Partha+2025, Biswas+2025}, many other mechanisms have been proposed. 

In fact, TTVs have also been observed in multi-planet systems \citep{Holman2010, Fabrycky2012, Masuda2013, Xie2013,
Jontof2015, Xie2014,
MacDonald2016, Mills2016}. These TTVs reconfirm the existence of additional planets in systems, and the measured TTV amplitudes help to improve the precision of planetary masses \citep{Lithwick2012, Masuda2014, Hadden2016, HaddenLithwick2016, Hadden2017}. 
Therefore, TTVs have become a useful technique to detect additional planets in known exoplanet systems \citep{Agol2005, Holman2005}.

Moreover, exoplanet resonant chains have been discovered and confirmed in many systems \citep{Mills2016, Gillon2017TRAPPIST,    Christiansen2018, MacDonald2021Kepler80, 
Leleu2021TOI178, MacDonald2022K2-138, Dai2023TOI1136, Masuda2024Kepler51}.
The convergent 
disk migration, an idea
similar to the drag-induced inward migration
proposed in \cite{JiangYeh2004}, has been considered as a possible mechanism to 
produce resonant pairs and chains in these exoplanet systems. If this mechanism did operate efficiently, many more exact resonant pairs and chains would have been observed.
However, according to the Kepler's observational results,
many near-resonance systems are actually deviated from 
exact resonances.
The statistic result  
that the number of resonant configurations decreases as a function of host star's age \citep{Dai+2024}
is consistent with a time-dependent deviation process of 
moving away from exact resonances. 
A few possibilities were proposed to address
this leaving from exact resonance, which might occur 
during convergent disk migrations \citep{Goldreich2014, Chen+2025, 
Charalambous2025}. 

On the other hand,
\cite{MacDonald2018} showed that more than one mechanisms 
could reproduce those resonant chains in their study, 
indicating 
that convergent disk migrations might not be the only choice. 
However, one shall be cautioned that only four compact systems with short orbital periods 
were included in \cite{MacDonald2018}.

Among these discoveries, a database produced by space-telescope survey projects such as Kepler or TESS could make an important
contribution and play a leading role \citep{Dai2023TOI1136}. 
The continuous monitoring nature of these projects makes it possible to have long baselines of many transit epochs for
both known and newly found transiting exoplanets.
These long baselines provide opportunities of detailed TTV analyses, which lead to
the measurements of physical and orbital parameters of known planets, and also the detection of 
many new planets \citep{Hoyer+2022, Wilson+2022, Korth+2024ApJ, Piotto+2024MNRAS,
Greklek+2025, Jiang+2025, Egger+2025}.  

Kepler-396 (KOI-2672) is a G-type star 
with a stellar mass of approximately 0.946 $M_\odot$ 
and a radius close to that of the Sun \citep{Berger2020II}.
Exoplanets Kepler-396 b and c were shown to exist and confirmed to be a pair near resonance through the
TTV technique \citep{Yang2013, Xie2014}. 
The orbital periods of Kepler-396 b and c 
are about 43 days and
88 days, respectively. Their periods are relatively large among resonant pairs of most exoplanet systems.
With larger orbital radii, 
it is unclear whether Kepler-396 b and c have experienced 
convergent disk migration during their evolution.
These multi-planet systems act as precious laboratories for testing theories of planetary formation and migration history.

Thus, the Kepler-396 system, with its two confirmed transiting planets, stands out as an important case to be investigated. Some important parameters provided by \cite{Jontof2022} are listed in
Table \ref{tab:origin-value}.
The Kepler-396 b and Kepler-396 c planetary radii are
estimated to be in the range of 3 to 6 $R_\oplus$, 
indicating that both planets could be mini-Neptunes. 
However, their masses 
have been reported differently with huge uncertainties in various papers. 
Though these two planets are close to 1:2 mean-motion resonance, the uncertainty in mass makes it 
difficult to determine the dynamical properties such as stability, chaos, and resonance of 
this planet pair. 
Later, both \cite{Battley2021} and
\cite{Jontof2022} combined TESS data with Kepler data and 
performed further TTV analyses. It was shown in \cite{Jontof2022} that there could be another planet in this system. Without a complete exploration of parameter space,
they proposed two possible orbital models for this third planet. 

Moreover, comparing Kepler-396 with the majority of other resonant planetary systems, the orbital periods of 
Kepler-396 b and c are relatively longer.
Given the possibility of the existence of a third planet,
it will be very interesting to investigate whether 
Kepler-396 is a resonant chain with longer orbital periods.

Therefore, our goal in this work is, through a new TTV analysis with all available Kepler and TESS data,
to obtain a comprehensive 
update of physical and orbital properties 
of Kepler-396 b and c, examine the existence 
and the orbit of an additional planet, and check 
whether it is a three-planet resonant chain.  
For the rest of sections, We first briefly mention previous observations in Section \ref{sec:obser}.  The new TESS data and the 
corresponding light-curve analysis will be described in 
Section \ref{sec:newTESS}.
The analysis of transit timing variations with two-planet and 
three-planet models will be given in Section \ref{sec:ttv}.
We then determine planet properties such as planet radii
and planet density in Section \ref{sec:pla_pro}.
In Section \ref{sec:orbits}, the planetary orbits are simulated and the resonances are investigated. 
We provide conclusions in Section \ref{sec:conclusions}.

\section{Previous Kepler and TESS Observations} 
\label{sec:obser}
The transit events of the star Kepler-396 (KOI-2672) were first captured by Kepler Space Telescope \citep{Holczer2016, Morton2016, Thompson2018}. In ensuing years,
Kepler-396 was also monitored by TESS in Sectors 14, 15, 41.
Thus, both \cite{Battley2021} and \cite{Jontof2022} were able to include TESS data into their TTV analyses. 
Among these two, \cite{Jontof2022} employed the 
Pixel-Level-Decorrelation (PLD) method to produce light curves from flux data. The determined mid-transit times for the transit events in Sector 14, 15, 41
were presented and provided in their paper.

\section{New TESS Observations and Light Curve Analyses}
\label{sec:newTESS}

Following the 
previous Kepler and TESS observations mentioned in Section \ref{sec:obser}, TESS performed further observations on Kepler-396 in Sectors 54, 74, 76. Here 
we take advantage of these new observations, i.e.
using the data to obtain 
light curves and corresponding mid-transit times. This would 
extend the TTV baseline, and ultimately lead to stronger constraints 
on the architecture of Kepler-396 planetary system.
The Pre-Search Data Conditioning Simple Aperture Photometry (PDCSAP) flux \citep{smith2017,smith_pdc2017}, calibrated through the Science Processing Operation Center (SPOC) pipeline \citep{Jenkins2016}, would be used here.  
The two-minute light curves were downloaded from the Mikulski Archive for Space Telescopes (MAST).

For these new TESS data in Sectors 54, 74, 76, we employ the
transit search package {\tt \string nuance}
\citep{Garcia+2024AJ} to search all transit signals of Kepler-396 planets. All available TESS light curves in Sectors 54, 74 and 76 are examined. 
The result is that all planet transits predicted by the ephemeris derived in \cite{Thompson2018} are identified
and validated. For the planet Kepler-396 b, there is
one transit in Sector 74 and another one in Sector 76.
For Kepler-396 c, there is one transit in Sector 54
and another one in Sector 74. 
It turns out that these two 
planets' transits in Sector 74 overlap and need a special
treatment for the light-curve fitting.

\begin{figure*}
\centering
\includegraphics[width=0.95\textwidth]{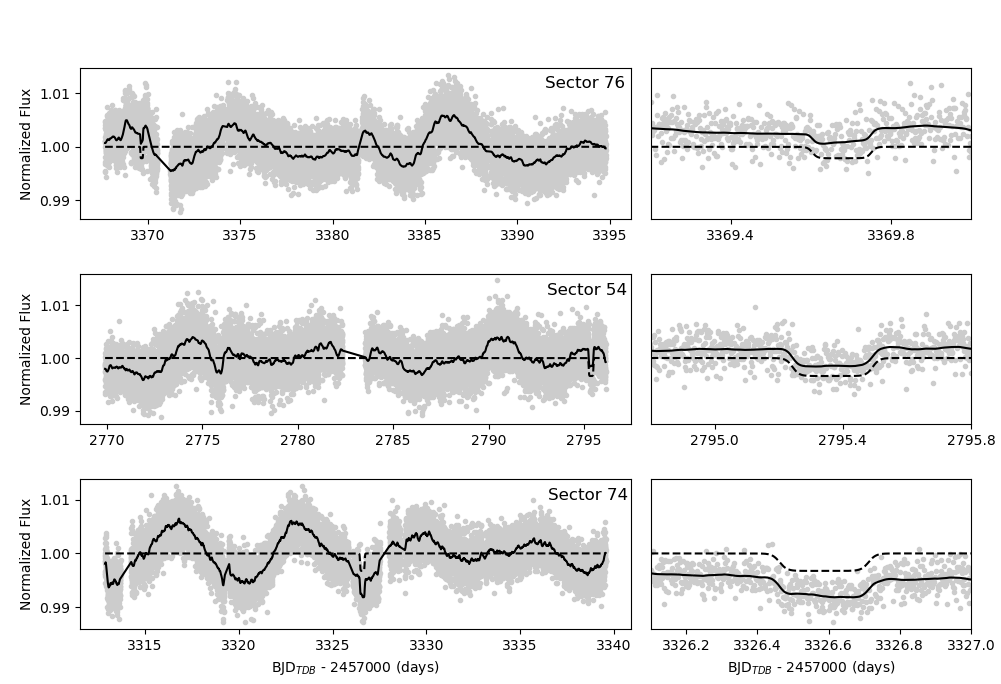}
\caption{The global light curves of Kepler-396 in TESS Sectors 76, 54, 74. The gray points are TESS data, the solid curves are the detrended global light curves, and the dotted lines indicate where the transit signals are. Right panels are the zoomed-in versions of left panels near the transit evens.}
\label{fig1_nuance}
\end{figure*}

\begin{figure*}
\centering
\includegraphics[width=0.95\textwidth]{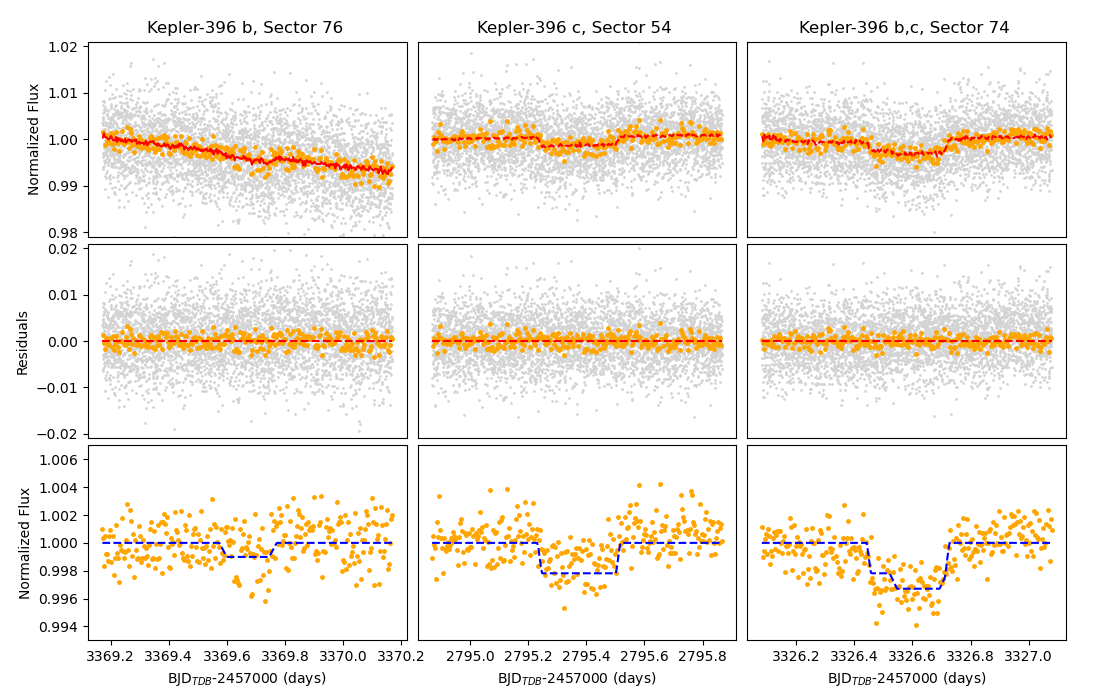}
\caption{The TESS transit light curves of Kepler-396 b and c. In each panel, the gray points are the TESS flux data and 
the orange points are 5-minute binned data. In top panels,
the red curves are the best-fit PLD models. The middle panels are the residuals for the best-fit PLD models. The blue lines
in the bottom panels are the trapezoid part of the best-fit 
PLD models.}
\label{figlc_bc}
\end{figure*}

The results are presented in 
Figure \ref{fig1_nuance}, in which the gray points are TESS data, the solid curves are the detrended global light curves, and the dotted lines
indicate where the transits are. To make the transit signals clear, 
the zoomed-in versions focusing on the transit signals are provided in the right panels of Figure \ref{fig1_nuance}. 
In order to proceed the PLD analysis later, 
we need to choose an aperture from the 
Target-Pixel-File (TPF).  
To show our aperture here,
the first image of Kepler-396 in Sector 74, downloaded through {\tt \string lightkurve} package, is presented in Figure \ref{fig_tpf} (See Appendix A). In order to avoid the contamination from nearby stars, seven pixels marked by 
red slashes shown in Figure \ref{fig_tpf} are chosen in this work. Centering on the mid-transit times determined
by the transit detection package {\tt \string nuance}, 
pixel-level light curves with a length of one day are
thus obtained for Sector 54, 74, and 76. 
The final normalized light curves are presented as the gray points in top panels of Figure \ref{figlc_bc}. 

\begin{table*}
\caption{
Some essential stellar and planetary parameters, where $T_{14}$ and $T_{23}$ are transit parameters (see the main text for details).}
\label{tab:origin-value}
\begin{center}
\hspace{-2.0cm}
\begin{tabular}{|c|c|c|c|c|}\hline
Star & surface gravity ${\rm log}$ g (cgs) & mass ($M_\sun$)& effective temperature (K)  & metallicity [Fe/H]  \\ \hline
Kepler-396  & 4.47 & 0.946 & 5384 & -0.129
         \\ \hline\hline
Planet &orbital period (days)&$T_{14}$ (hours) & $T_{23}/T_{14}$ &inclination (degree)\\ \hline
b& 42.992 & 4.73 $\pm$ 0.11 & 0.837 $\pm$ 0.233& 89.34   \\ \hline
c & 88.512& 6.82 $\pm$ 0.1 & 0.9 $\pm$ 0.013& 89.86 
\\ \hline  
\end{tabular}
\end{center}
\end{table*}

\begin{table*}
\caption{The PLD fitting process and retrieved results, where
$U()$ and $N()$ indicate uniform and normal distributions respectively
(see the main text for parameter definitions).}
\label{tab:PLDPrior}
\hspace{-0.5cm}
\begin{tabular}{|l|l|c|c|}\hline
Sector, Planet         & Parameter &  Prior Distribution & Retrieved Values \\ \hline
 76, Kepler-396 b    
 & $T^{b}_{mid}$ (BJD$_{\rm TDB}$ - 2457000) &  $U$(3369.57, 3369.77)   &  3369.6727 $\pm$ 0.0288  \\
& $T^{b}_{14}$ (hours)  &     $N(4.73, 0.11)$  & 4.73 $\pm$ 0.11  \\
& $f^{b}_T$      & $N(0.802, 0.099)$  &  0.741 $\pm$ 0.183\\
& Transit Depth $D^{b}$ (ppm) & $U$(0, 5000)  &   1010$\pm$ 724         \\  \hline\hline
 54, Kepler-396 c       & $T^{c}_{mid}$ (BJD$_{\rm TDB}$ - 2457000)  &  $U$(2795.27, 2795.47)  & 2795.3749 $\pm$ 0.0221 \\
& $T^{c}_{14}$ (hours)  &   $N(6.82, 0.1)$ &  6.82 $\pm$ 0.1        \\
& $f^{c}_T$      &  $N(0.9, 0.013)$ &  0.899$\pm$ 0.013      \\
& Transit Depth $D^{c}$ (ppm) & $U$(0, 7000)  & 2184 $\pm$  841     \\  \hline \hline
 74, Kepler-396 b,c       & $T^b_{mid}$ (BJD$_{\rm TDB}$ - 2457000)  & $U$(3326.52, 3326.72) &  $T^b_{mid}$=3326.6200 $\pm$ 0.0582   \\
& $T^c_{mid}$ (BJD$_{\rm TDB}$ - 2457000) & $U$(3326.52, 3326.72) & $T^c_{mid}$=3326.5860 $\pm$ 0.0286      \\
& $T^b_{14}$ and  $T^c_{14}$ (hours)  & Fixed  &   $T^b_{14}$=4.73, $T^c_{14}=6.82$   \\
& $f^b_T$ and $f^c_T$     &  Fixed  &  $f^b_T=0.741$, $f^c_T=0.899$   \\
& Transit Depth $D^b$ and $D^c$ (ppm) & Fixed &   $D^b$=1010 ,$D^c$=2184             \\  \hline 
\end{tabular}
\end{table*}

With the above three TPF transit light curves, following \cite{Deming2015PLD} and \cite{Jontof2022}, the PLD analysis is employed to obtain the transit models. Once the transit models are settled, the corresponding mid-transit times    
will be determined and used for all the rest related analysis in this paper.

The PLD analysis needs a trapezoidal transit model and a 
corresponding noise model.
For a trapezoidal transit model,
we set the time between the first and fourth contacts 
to be the "transit duration" $T_{14}$, and 
the time between the second and the third contacts 
to be the "bottom duration" $T_{23}$.
The fitting parameters of a trapezoidal transit model include the mid-transit time $T_{mid}$, the transit duration $T_{14}$,
the ratio $f_T\equiv T_{23}/T_{14}$, the transit depth $D$. 
Their reference values are shown in 
Table \ref{tab:origin-value}.
Later, the superscripts of these parameters would represent planet b and c respectively. 
The corresponding noise model $M^{t}_{PLD}$ is obtained as 
\begin{equation}
M^{t}_{PLD}=c_0(t-t_0)+\frac{\sum_{i=1}^{7} c_i P_i^t}{\sum^7_{i=1}P_i^t}, \label{eq:M_PLD}
\end{equation}
where $c_0$ is the coefficient of a 
linear function; $t_0$ is the starting time of the selected time interval; $P_i^t$ is the pixel value of 
a chosen $i$-th pixel 
at a particular time $t$ and $c_i$ is the PLD coefficients of the same pixel.

The trapezoidal transit model and 
the corresponding noise model are added to become a combined
transit light-curve model, which would be taken to fit our light curves. The fitting parameters 
include those in trapezoidal transit model and all
PLD coefficients $c_i$.The joint posterior distribution was sampled using the {\tt \string dynesty} package, the MIT-licensed Dynamic Nested Sampling package (\cite{Speagle2020MNRAS},\cite{Skilling2004AIPC},\cite{Higson+2019}) for estimating Bayesian posteriors and evidence. 
Similar to those ensemble sampling methods such as Markov Chain Monte Carlo (MCMC), the behavior of Nested Sampling can be sensitive to the number of live points used. Increasing the number of live points results in smaller changes in the prior volume over time, thereby improving the effective resolution but also increasing the computational runtime. Here, 20000 live points are used for each light curve.
Thus, through the above process, the transits in Sector 76 and 54 are fitted 
by a trapezoidal transit model with four parameters
separately. However, in Sector 74, two transit events occur around the same time and these transits overlap.  
Therefore, a two-trapezoid transit model, in which six parameters,$T^{b}_{14},T^{c}_{14},f^{b}_{T},f^{c}_{T},D^{b},D^{c}$ are fixed to be the best-fit values from
the previous PLD transit fitting processes in Sector 76 and 54, is employed to fit the overlapping light curve and obtain
two more mid-transit times $T^b_{mid}$ and $T^c_{mid}$. 

In Table \ref{tab:PLDPrior}, the PLD fitting prior distributions of all related parameters are listed. We allowed the coefficient $c_i$ to be between -5 and 5. 
The retrieved results presented in Table \ref{tab:PLDPrior} and the values of transit depth $D^b$ and $D^c$ are consistent with \citep{Lissauer+2024PSJ} ($1110.7\pm 40.7$ ppm and $2522.9\pm 47.2$ ppm) for planet b and planet c respectively. In Figure \ref{figlc_bc}, the data and best-fit models for a transit of Kepler-396 b in Sectors 76; transits of Kepler-396 c in Sectors 54 and a composite model of planet b and planet c in Sector 74 are also presented.
In each upper panel of Figure \ref{figlc_bc}, the gray points represent the normalized TPF light-curve data; the orange dotted points indicate the five-minute binned data, and the red line shows the best-fit trapezoidal transit with a noise model. The middle panels display the corresponding residuals. After subtracting the linear trend,
the five-minute binned data are shown as orange points and  
the best-fit trapezoidal transit model are shown as blue lines in the bottom panels.

\section{Transit Timing Variations: Two-Planet Model and Three-Planet Model}
\label{sec:ttv}

Exoplanets Kepler-396 b and c were first discovered from Kepler data through the TTV technique \citep{Yang2013, Xie2014}. 
Combining TESS data with Kepler data to form
a longer baseline, \cite{Jontof2022} obtained a result with larger TTVs and pointed out the possibility of an additional planet. To advance further, here we examine this possibility through considering later new TESS data
and a complete parameter exploration of TTV modeling. 
In Table \ref{tab:Planet b} and Table \ref{tab:Planet c} of Appendix B, we list all the mid-transit times for planets Kepler-396 b and c respectively, including those derived from new TESS data as described in Section \ref{sec:newTESS}.

A convincing model shall fit the transit timings 
of both Kepler-396 b and c well. It was previously shown in \cite{Battley2021} that the two-planet model cannot produce a TTV model being consistent with the transit timing of Kepler-396 c.
Therefore, in order to make theoretical TTV curves have a better fitting to observational TTVs, 
\cite{Jontof2022} proposed two different 
three-planet models. With an updated longer baseline TTV data, here we would systematically explore all possible parameter spaces
and search for the best-fit three-planet model. 
Since orbital period is one of the key parameters affecting
TTV, we search for the orbital period of the third planet from three different intervals. The first one is exterior to two known planets and ranges from 90 days to 180 days
including Model 3pA1 and Model 3pA2 with different prior distributions of the inclination of planet d.
The second one is between two known planets and ranges from 45 days to 88 days (Model 3pB). The third one is interior to two known planets and ranges from 5 days to 40 days (Model 3pC). 

Thus, here we employ the package {\tt \string TTVFast}  \citep{Deck2014} to produce theoretical TTV signals and fit to the observational counterparts for both two-planet and three-planet models. Each planet in the system  was fitted for parameters mass $m_i$, orbital period $p_i$, orbital eccentricity $e_i$,  argument of pericenter $w_i$  (via $ec\equiv e\cos w_i$ and $es\equiv e\sin w_i$), and mean anomaly $MA_{i}$,
where $i = b, c, d$ as the sub-indexes of variables associated to the planet Kepler-396 b, Kepler-396 c, and the possible Kepler-396 d, respectively. 
Finally, the orbital inclinations of planet b, c are fixed
but we have two different settings for the 
the inclination of planet d.
In Model 3pA1, 3pB, and 3pC,  
this new planet's orbital inclination $i$ is a fitting parameter and its value is set to be uniformly distributed between 0 and 90 degree. It
is allowed to vary during the fitting process as it is desired to see its best-fit value. In contrast, 
to test what the outcome could be when this new planet is 
coplanar with planet c, the inclination of planet d is fixed
to be the same as the value of planet c in Model 3pA2.

\begin{table*}
\caption{Initial parameter settings of the Nelder-Mead fitting process. The intervals and prior distributions of initial values of fitting parameters are listed here,
where $U()$ indicates a uniform distribution. 
Note that the inclinations of planet b, c are always 
fixed to be initial values, i.e. not a fitting parameter,
for all models. As for the inclination of planet d,
it is treated as a fitting parameter for 
Model 3pA1, Model 3pB, and Model 3pC, but a 
fixed parameter for Model 3pA2. See the main text for parameter definitions.
}
\label{table:NMini}
\hspace{-1.9cm}
\begin{tabular}{|l|c|c|ccccc|}\hline
Model &Planet &$m(M_\earth)$&  $p$ (days)& $e\cos w$ & $e\sin w$  &$MA$(degrees) &  $i$(degrees) \\ \hline
& b & $U$(8.514, 10.595) & 
$U$(42.9919, 42.9923)& $U$(-0.01, -0.005)& 
$U$(-0.007, -0.003) & $U$(0, 360) & 89.34  \\
2p&  c &  $U$(10.406, 13.528)& 
$U$(88.5163, 88.5173)&$U$(0.053, 0.065)& $U$(0.018, 0.026)& $U$(0, 360) & 89.86         \\ \hline
 &  b & $U$(4.12, 8.37) & $U$(42.9924, 42.9928) &
$U$(-0.127, -0.07) &$U$(-0.039, -0.025) & $U$(0, 360) & 89.34 \\ 
3pA1 & c &$U$(3.53, 6.18) &
$U$(88.5216, 88.5256) &$U$(-0.056, -0.019) &
$U$(-0.084, -0.054) & $U$(0, 360) & 89.86 \\
3pA2& d &$U$(5.58, 5.77)&$U$(90, 180) &
$U$(-0.12, -0.056)& $U$(-0.062, 0.085) & $U$(0, 360)&
89.86
\\  \hline
  & b & $U$(5.72, 7.43) & $U$(42.9917, 42.9921) &
$U$(-0.072, -0.053) & $U$(0.015 ,0.024) & $U$(0, 360) 
& 89.34 \\ 
3pB & c &$U$(4.52, 5.61) &
$U$(88.5167, 88.5179) &$U$(-0.037, -0.005) 
& $U$(0.059, 0.075) & $U$(0, 360) & 89.86 \\ 
&  d & $U$(0.16, 0.22)&$U$(45, 88) &$U$(0.012, 0.026)&
$U$(-0.165, -0.144) & $U$(0, 360) & 89.86
\\  \hline
 & b & $U$(5.72, 7.43) & $U$(42.9917, 42.9921) &
$U$(-0.072, -0.053) & $U$(0.015, 0.024) & $U$(0, 360) & 89.34 \\ 
3pC & c & $U$(4.52, 5.61) &
$U$(88.5167, 88.5179) & $U$(-0.037, -0.005)
& $U$(0.059, 0.075) & $U$(0, 360) & 89.86\\ 
&  d &$U$(0.16, 0.22)&$U$(5, 40) &$U$(0.012, 0.026)&
$U$(-0.165, -0.144) & $U$(0, 360) & 89.86  
\\  \hline
\end{tabular}
\end{table*}

\begin{table*}
\caption{The prior distributions of MCMC fitting parameters, where $U()$ and $N()$ indicate uniform and normal distributions respectively, and $MA^{NM}$ is the solution derived from the Nelder-Mead algorithm (see the main text). }
\label{table:Bnd}
\hspace{-2.5cm}
\begin{tabular}{|c|c|c|c|c|c|c|c|}\hline
Planet &Model &  $p$ (days) &$i$ (degrees) &$m$ $(M_\earth)$& $e\cos w$ & $e\sin w$  & $MA$ (degrees) \\ \hline
b & 2p,3pA1,3pA2,3pB,3pC & $U$(42.98,43.02)& 89.34 (fixed)  & & &  & \\ \cline{1-4}
c& 2p,3pA1,3pA2,3pB,3pC &$U$(88.48,88.54)&89.86 (fixed) &&  & & \\\cline{1-4}
 & 3pA1 &  $U$(90,180) & $U$(0,90) &$U$(0.1,30)&$U$(-1,1) &$U$(-1,1)  & $N(MA^{NM},1)$
 \\  \cline{2-4}
 &3pA2 & U(90,180)  &  89.86 (fixed) & & & & \\ \cline{2-4}
d & 3pB &$U$(45,88) & $U$(0,90) &  & &  &\\ \cline{2-3}
& 3pC & $U$(5,40) & &  & &  &\\  \hline
\end{tabular}
\end{table*}

Following the process in \cite{Jiang+2025}, 
a best-fit theoretical TTV
solution was first searched through the Nelder-Mead algorithm \citep{GaoHan2010}, which is embedded in the package {\tt \string SciPy} \citep{Virtanen2020}, from the suggested parameter intervals 
in \cite{Jontof2022}. 
The setting of initial values of parameters of all 
different models of the above numerical searching are presented in Table \ref{table:NMini}.
This searching process is repeated for 100 times and the one with smallest chi-square value is picked as the 
best-fit solution at this first fitting stage.
Then, using that best-fit solution of Nelder-Mead algorithm as a basis to set initial values for all fitting parameters, MCMC sampling was later performed through {\tt \string emcee} package
\citep{Foreman2013} in order to further refine the solution and obtain the uncertainties. During the MCMC sampling, 32 walkers were initialized in a small ball around the initial parameters and each walker was run for 50000 steps.  In general, there are two popular ways of move for these walkers. One is the differential evolution MCMC  \citep{Nelson2014}, the other is a snooker proposal using differential evolution \citep{Braak2008}.
The weighted mixture of these two moves, selecting either a differential evolution MCMC with 80 $\%$ probability or a snooker proposal using differential evolution with 20 $\%$ probability, is used at each step.  
With the prior distributions of the fitting parameters in Table \ref{table:Bnd}, the MCMC fitting process was performed. 

Then, the MAP (maximum a posterior) solution is set to be our result of the MCMC fitting process \citep{Nasci2023, Nasci2024}, and the standard deviations of the posterior distributions of the parameters are used as uncertainties. Table \ref{table:bestMCMC} gives the values of the best-fit parameters and the corresponding uncertainties of the above MCMC sampling process of all models. 

\begin{figure*}
\centering
\includegraphics[width=0.85\textwidth]{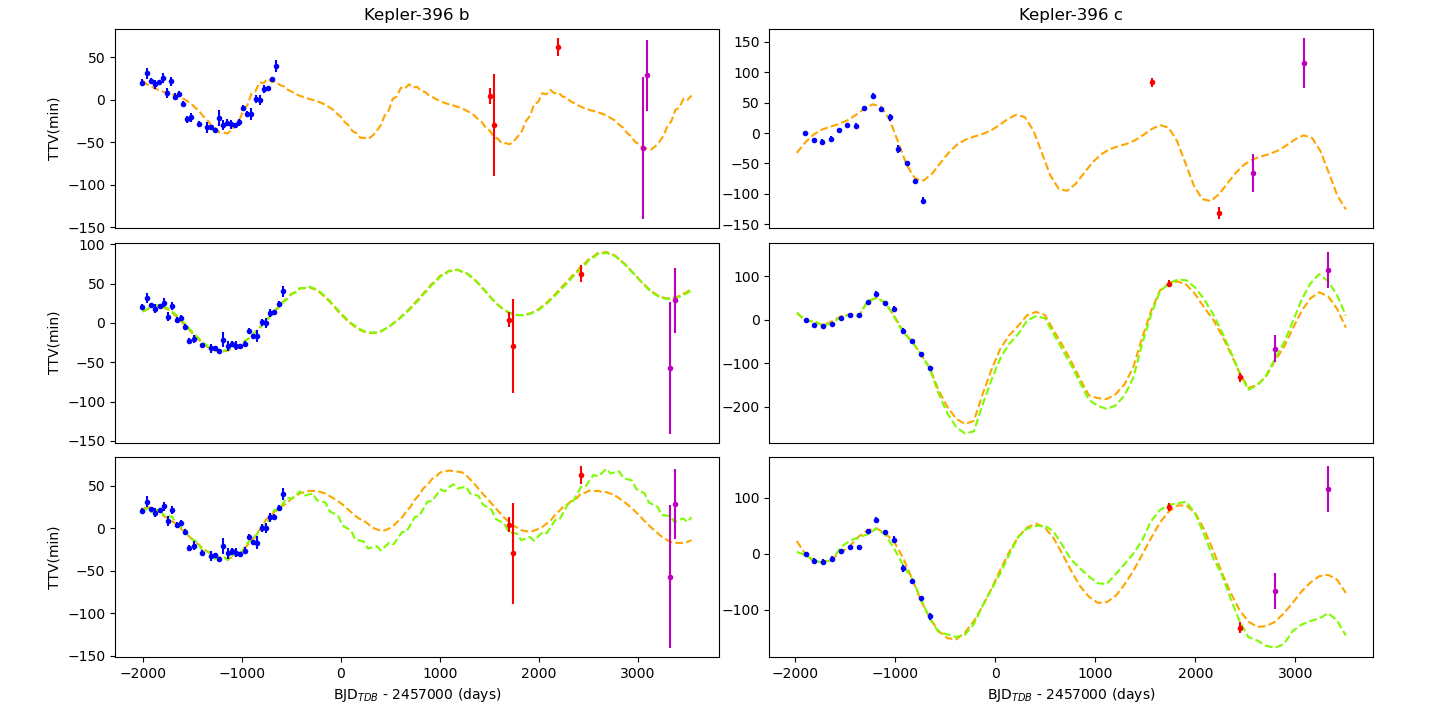}
\caption{The TTV plots for Kepler-396 b (left panels) and 
Kepler-396 c (right panels). In each panel, points with error bars are observational mid-transit times from Table \ref{tab:Planet b} and Table \ref{tab:Planet c} 
of Appendix B, and curves are theoretical TTV models. Blue points are the Kepler data from \cite{Thompson2018}, red points are the TESS data adopted from \cite{Jontof2022}, and purple points are our results.
The orange curves in top panels are for two-planet model (Model 2p). In middle panels, the orange curves and the green curves are for Model 3pA1 and Model 3pA2 respectively. In bottom panels, the orange curves are for Model 3pB, and the green curves are for Model 3pC.}
\label{fig_ttv}
\end{figure*}

Here we first present the two-planet model, i.e. Model 2p, 
in upper panels of Figure \ref{fig_ttv}. The blue points are derived from Kepler data \citep{Thompson2018}, red points are from TESS data \citep{Jontof2022}, and purple points are obtained through our new TESS data. We find that some red points are far from the theoretical TTV curve.
These results are similar to those shown in \cite{Jontof2022}.
In addition, the theoretical TTV curves of three-planet models, i.e. Model 3pA1, Model 3pA2, are presented in 
middle panels of Figure \ref{fig_ttv}. It is clear that
these curves are closer to all observational points.
The results of Model 3pB and 3pC are shown in
bottom panels of Figure \ref{fig_ttv}. 
The posterior distributions of the parameters of MCMC samples 
from Model 3pA1 are also shown in Figure \ref{fig:corner_Ax} 
of Appendix A as an example, and it is clear that some parameters are slightly correlated.

We then calculate the reduced chi-square values $\chi^2_{red}$ and Bayesian Information Criterion (BIC) values for best-fit solutions from all models, which is
summarized in Table \ref{tab:chi}. 
It shows that the $\chi^2_{red}$ and BIC values 
become much smaller after adding the third planet. 
In this table, it is clear that Model 3pA1 and Model 3pA2 
lead to best-fit solutions with
reduced chi-square value 2.24 and 2.34 respectively,
so they are both good fittings superior to other models. 

After ruling out the two-body model, which has a reduced chi-square value 12.97,
here we investigate the orbital stability of best-fit solutions of all three-body
models through orbital simulations. 
Our simulations show that the least massive planets in Model 3pB and Model 3pC 
are scattered out after several dynamical times, indicating these are unstable systems.
In contrast, both Model 3pA1 and Model 3pA2 are stable for $10^6$ years.
Considering that Model 3pB and Model 3pC are unstable and
present poorer fitting in TTV plots shown in the bottom panels of Figure \ref{fig_ttv}, 
Model 3pA1 and Model 3pA2 are left as the worth-considered solutions derived from our updated transit timing baseline. From the above simulations, we can see that the relative inclinations 
between any pairs of three planets in Model 3pA1 is about 10 degrees (see further details
in Section 6), which would significantly reduce the time of co-transit between planet b and c.
In Model 3pA2, the relative inclinations are less than one degree and
three planets are almost coplanar, so the transit event of the third planet shall be detectable.
In order to search for the transit of this third planet, i.e. Kepler-396 d, from Kepler data, we revisited all Kepler 
data and used {\tt\string nuance} package to seek for the transit of this third planet from possible windows. Finally, we failed to detect any transit event of the third planet.

To summarize, on one hand, the fact that planet b and c are currently co-transit planets makes Model 3pA1 less likely. On the other hand, 
the failure to detect the transit event of planet d shows that Model 3pA2 might not be the
real solution. Nevertheless, since the values of best-fit parameters derived from both models are
very close except the inclinations, we decide to regard them equally as best-fit solutions
hereafter. With different choices of prior, we conclude that 
the current observational data can only provide these best-fit results with an uncertain inclination. 
The true solution with confirmed inclination could be the one
from this solution family. We leave it open for future work.

\begin{table*}
\caption{The MCMC best-fit solutions of two-planet and three-planet models. The values of planetary mass $m$, orbital period $p$, $ec$, $es$, mean anomaly $MA$, and orbital inclinations $i$ are listed.}
\label{table:bestMCMC}
\hspace{-3.05cm}
\begin{tabular}{|c|c|c|c|c|c|c|c|}\hline
Model &Planet &$m (M_\earth)$&  $p$ (days)& $e\cos w$ & 
$e\sin w$  &$MA$(degrees) & $i$ (degrees)\\ \hline
2p & b &13.47 $\pm$ 0.75 & 42.9869 $\pm$ 0.0003 &0.0060$\pm$ 0.0005 &-0.0085 $\pm$ 0.0008& 135.92 $\pm$ 2.32 &89.34 \\
 &c&  11.94 $\pm$ 0.75 & 88.5430 $\pm$ 0.0013 & 0.0903 $\pm$ 0.0033 & 0.0155 $\pm$ 0.0020 &  344.14 $\pm$ 1.19 &89.86\\ \hline
3pA1 & b & 8.89$\pm$ 0.38& 42.9942 $\pm$ 0.0001 &-0.1248 $\pm$ 0.0043& -0.0374$\pm$ 0.0021 &259.66 $\pm$ 1.00 &89.34\\ 
& c & 4.29$\pm$ 0.18 & 88.5085 $\pm$ 0.0011 & 
-0.0713  $\pm$ 0.0026 & -0.0813 $\pm$  0.0032 & 143.85 $\pm$ 0.82 & 89.86\\ 
&d & 4.66$\pm$ 0.82 & 144.0110$\pm$ 0.0243& 
-0.1455  $\pm$ 0.0059 & 0.0225 $\pm$  0.0058 & 152.54 $\pm$ 1.24 &76.71 $\pm$ 3.24\\ \hline  
3pA2 & b & 9.22$\pm$ 0.34& 42.9944 $\pm$ 0.0001 &-0.1269 $\pm$ 0.0046& -0.0349$\pm$ 0.0027 &261.17 $\pm$ 1.68 &89.34\\ 
& c & 4.18$\pm$ 0.21 & 88.5097 $\pm$ 0.0020 & 
-0.0725  $\pm$ 0.0027 & -0.0840 $\pm$  0.0034 & 143.57 $\pm$ 0.89 & 89.86\\ 
&d & 4.25$\pm$ 0.41 & 144.0372$\pm$ 0.0188& 
-0.147  $\pm$ 0.007 & 0.008 $\pm$  0.009 & 146.50 $\pm$ 1.12 & 89.86\\ \hline  
3pB & b & 10.69$\pm$ 0.72 & 42.9919$\pm$ 0.0005 &-0.0625 $\pm$ 0.0034& 0.0199$\pm$ 0.0011 & 286.33 $\pm$ 0.67 &89.34\\
&c& 6.10 $\pm$ 0.33 & 88.5060 $\pm$ 0.0009 & -0.0030$\pm$ 0.0004 &  0.0258 $\pm$ 0.0032 & 267.61 $\pm$ 0.41 & 89.86\\
&d& 0.18 $\pm$ 0.02 & 71.2290 $\pm$ 0.0722 & 0.1209 $\pm$ 0.0053 &-0.0827 $\pm$ 0.0060& 217.21 $\pm$ 1.82 & 88.99 $\pm$ 3.51\\ \hline
3pC & b & 19.16 $\pm$ 1.17 & 42.9936 $\pm$ 0.0003 & -0.0343 $\pm$ 0.0018 & 0.0093 $\pm$ 0.0011 & 280.68 $\pm$ 1.61 &89.34\\
&c & 1.16 $\pm $ 0.22 & 88.5224 $\pm$ 0.0016&  -0.0095 $\pm$ 0.0009 & 0.0223 $\pm$ 0.0018 & 252.07 $\pm$ 0.80 &89.86\\
&d& 2.16 $\pm$ 0.22 &32.2382 $\pm$ 0.0015 &-0.0005 $\pm$0.0079 & -0.4642 $\pm$ 0.0088 & 352.86 $\pm$ 1.19 & 87.52 $\pm$ 1.94 \\ \hline
\end{tabular}
\end{table*}

\begin{table*}
\caption{The measure of goodness for best-fit two-planet and three-planet models. The number of mid-transit time $n$, the numbers of fitting parameter $m$, the chi-square values $\chi^2$, the reduced chi-square values $\chi_{red}^2$, BIC, and BIC value differences from the value of Model 2p, i.e $\Delta$BIC are listed.}
\label{tab:chi}
\hspace{1.95cm}
\begin{tabular}{|c|c|c|c|c|c|c|}\hline
Model & $n$ & $m$ & $\chi^2$  &$\chi_{red}^2$ & BIC & $\Delta$BIC\\ \hline
2p& 56 &10 & 596.44 & 12.97 &  636.69 & 0\\ 
3pA1 & 56 &16 & 89.56 & 2.24 & 153.97 &-482.73\\
3pA2 & 56 &15 & 95.96 & 2.34 & 156.34 &-480.35\\
3pB & 56 & 16 &151.79 & 3.79 &  216.20&-420.50\\
3pC &56 & 16 & 165.92 & 4.15&  230.33 &-406.37\\ \hline
\end{tabular}
\end{table*}

\section{Planetary Properties}
\label{sec:pla_pro}

Through the TTV analysis and fitting, in addition to 
investigating
the existence, the mass, and the orbit of 
a new planet Kepler-396 d,
we also obtain the most updated masses 
of Kepler-396 b and c. 

\begin{figure*}
\centering
\includegraphics[width=1.0\textwidth]{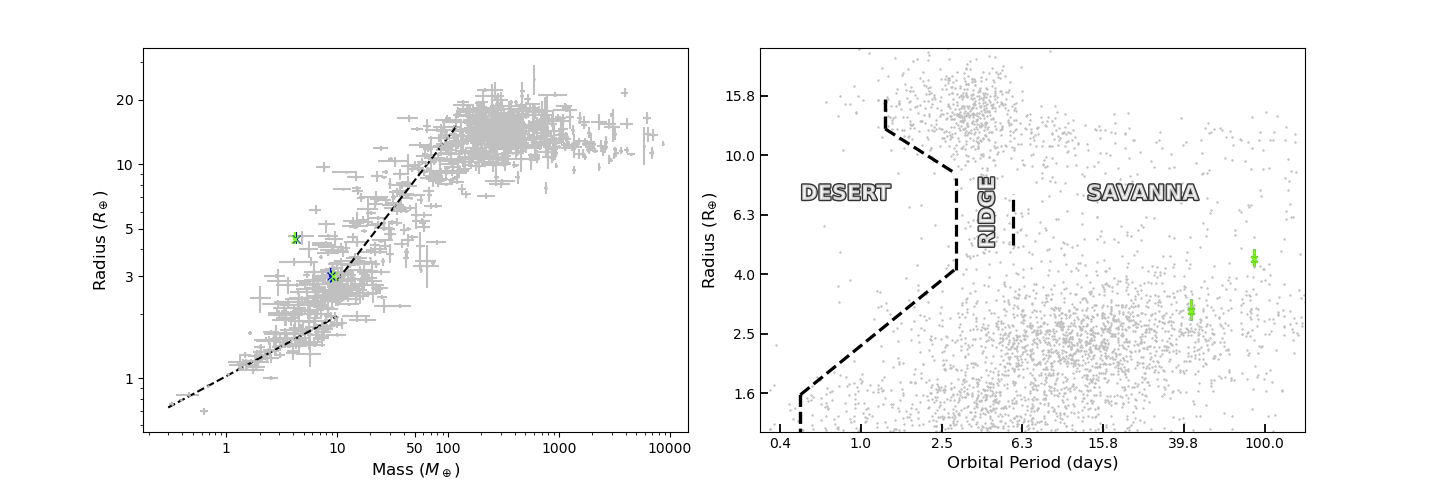}
\caption{The properties of Kepler-396 b and c. 
The left panel: Kepler-396 b and c (blue points for Model 3pA1 and green points for Model 3pA2) on the mass-radius plane, where gray dots are the planets 
from NASA Exoplanet Archive (https://exoplanetarchive.ipac.caltech.edu/, \cite{Christiansen2025}) with planetary radius and mass uncertainties below 30$\%$. The dashed lines are adopted from
\cite{Parc2024}.
The right panel: Kepler-396 b and c (blue points for Model 3pA1 and green points for Model 3pA2) on the period-radius
plane of the observed planet population with exo-Neptunian boundaries from \cite{Castro2024}.}
\label{fig_MR}
\end{figure*}

\begin{figure*}
\centering
\includegraphics[width=1.0\textwidth]{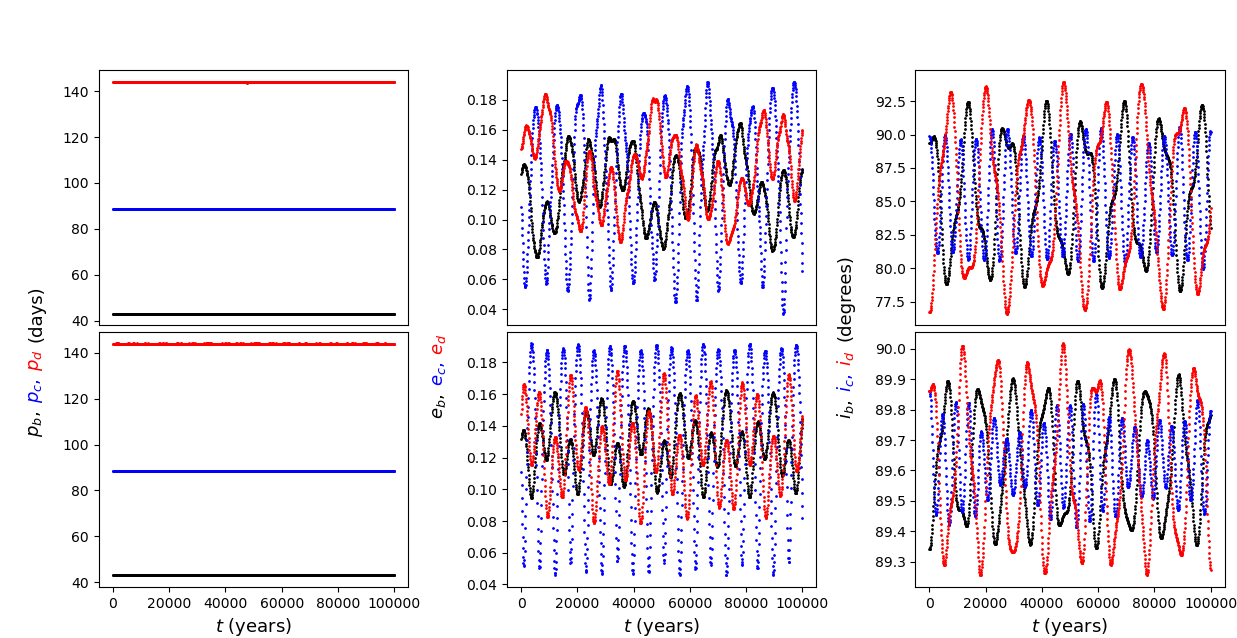}
\caption{The evolutions of period, eccentricities and inclination of the best-fit solution in Model 3pA1 and Model 3pA2  over $10^5$ years. The top panels are for Model 3pA1 and the bottom panels are for Model 3pA2 where 
black, blue, and red curves are for Kepler-396 b, c, d, respectively.}
\label{fig_bcx}
\end{figure*}

\begin{figure*}
\centering
\includegraphics[width=0.9\textwidth]{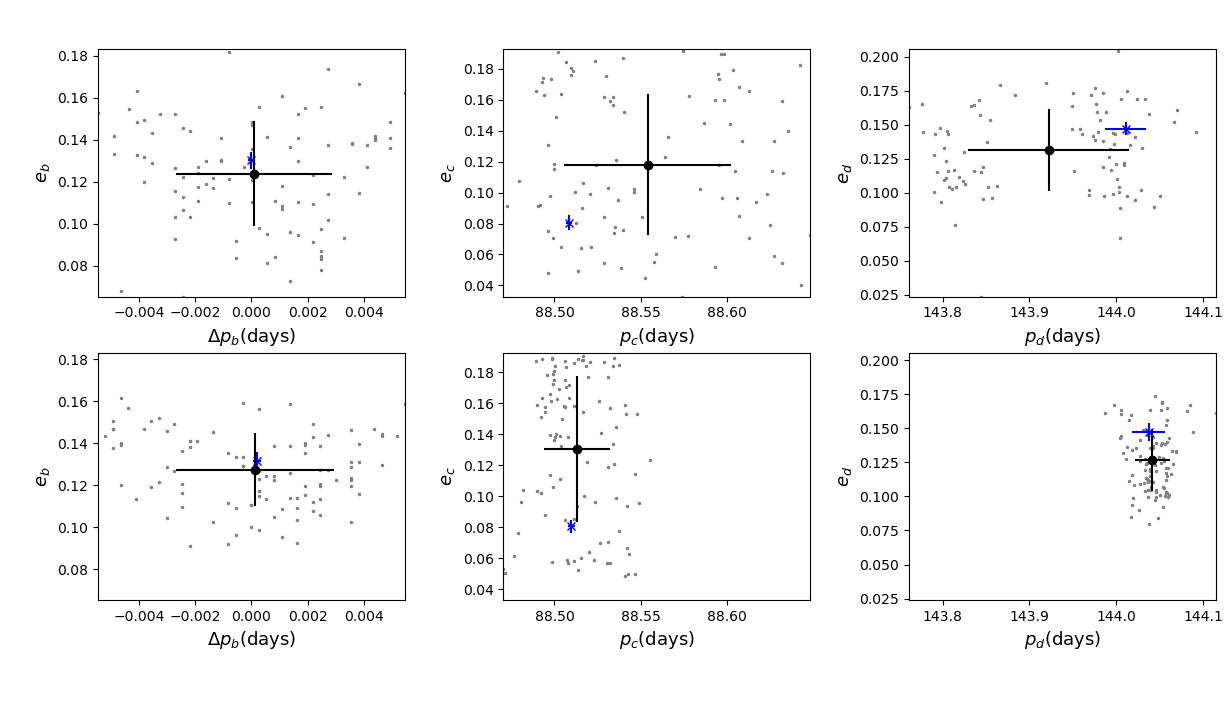}
\caption{The distributions on period-eccentricity plane at the end time $t=10^6$ years for 100 simulations in Model 3pA1 (top panels) and another 100 simulations 
in Model 3pA2 (bottom panels). Left panels are for planet b, 
middle panels are for planet c, and right panels are for planet d. For convenice, $\Delta p_b\equiv p_b-42.99423$. 
The gray points are the values of the periods and eccentricities at the time 
$t = 10^6$. In addition, the black points with
error bars present the means and standard deviations of the gray points, while the blue points with error bars show the
best-fit solutions and its associated uncertainties.
}
\label{figpe_final}
\end{figure*}

According to \cite{Lissauer+2024PSJ}, the values of planetary radii expressed as the unit of stellar radius,
i.e. $rp/rs$, for Kepler-396 b and c are $0.030487^{+0.001614}_{-0.000986}$ and 
$0.045546^{+0.001621}_{-0.000695}$, respectively. 
With the value $r_s$ being set as 0.903 
$R_\odot$ which is adopted from \cite{Berger2018},   
the planetary radii $r_p$ are determined to be 3.00 and 4.48 for planet b and c respectively.
Based on the planetary mass and radius we obtained, 
Kepler-396 c is confirmed to be a super-puff.

In fact, the existence of super-puff remains to be a puzzle.
There were a few attempts to explain it by proposed theoretical models.
It was suggested that planetary cores could acquire huge envelopes only if planets were located in a cool and low-density region of the proto-planetary disk. These models then proposed that super-puff formed farther
out and migrated inward \citep{Lee2016}. Others claimed that large internal 
heat fluxes might inflate planetary radii through tidal or 
ohmic dissipation \citep{Pu2017, Millholland2019}.
No matter how super-puffs gained their envelopes, they could have been stripped off due to photo-evaporation during their formation and evolution. Therefore, a mechanism which could replenish the accreted material 
could be an appealing choice. 

The relationship between an exoplanet’s radius and mass provides valuable constraints on its envelope fraction and internal structure. Therefore, studying the mass–radius relation is essential for understanding planetary composition and evolution.
In the left panel of Figure \ref{fig_MR}, 
Kepler-396 b is located near the boundary between the rocky and volatile-rich populations but clearly belong to the latter. In particular, the density of Kepler-396 c is lower than that of many volatile-rich planets. This intriguing result may provide valuable clues about their formation histories.

In the right panel of Figure \ref{fig_MR}, Kepler-396 b and Kepler-396 c lie in the Neptune savanna \citep{Bourrier2023}, a populated region in the period–radius diagram located at orbital periods longer than those defining the Neptune desert \citep{Mazeh2016A&A} and the Neptunian ridge \citep{Castro2024}. Both planets exhibit low densities as shown in Figure \ref{fig_MR},
consistent with expectations for planets residing in the savanna \citep{Castro2024}.

\section{The Orbital Stability and Resonances}
\label{sec:orbits}

In this section, dynamical simulations are performed 
through the orbit-integration code 
{\it REBOUND} \citep{ReinLiu2012} 
in order to investigate the 
orbital stability\citep{Petrovich2015, Kane2015,
Veras2017,  Liu2020,  Kovacs2022, Teixeira2023, 
Livesey2024}, regularity\citep{Volk2020, 
Hussain2020, Shevchenko2022}, and the resonances between Kepler-396 planets.

Here, three planets of Model 3pA1 and Model 3pA2 will be included in our simulations and the planetary masses and orbital elements of Kepler-396 b, c and d are set to be the values in Table \ref{table:bestMCMC}. 
In Figure \ref{fig_bcx}, the evolution of orbital periods, eccentricities and inclinations are presented for different planets in the simulation with an initial condition set as the best-fit solution in Table \ref{table:bestMCMC}.
The top panels and bottom panels of Figure \ref{fig_bcx} are for Model 3pA1 and 3pA2 respectively. The left panels of Figure \ref{fig_bcx} illustrate the evolution of orbital
periods for three planets, the middle panels illustrate the evolution of their eccentricities, and the right panels show the evolution of their inclinations for $10^5$ years. The periods remain close to their initial values and the eccentricities exhibit oscillations around their starting values. On the right panels, the inclination of planet d exhibits larger fluctuations, ranging from 77.5 degrees to 92.5 degrees in the top panel for Model 3pA1, compared with 89.3 degrees to 90.0 degrees in the bottom panel for Model 3pA2. The properties of Model 3pA1 and Model 3pA2 are similar except the range of inclinations. 
The future orbits of these three planets derived from the best-fit solutions appear dynamically stable. 

Furthermore,
to ensure a sufficient number of representative cases, 100 different initial conditions are considered, and their corresponding results would be obtained accordingly. The 100 sets of planetary masses and orbital parameters with the smaller chi-square values from the MCMC samples of the fitting process of Model 3pA1 and another 100 sets from Model 3pA2 are employed. During the simulations, the planetary masses are fixed, yet the orbital parameters are used as initial values and evolve as the function of time. 
These 200 dynamical simulations start from a time $t=0$ to a time $t=10^6$ years,
and the final snapshots from all simulations are shown on the period–eccentricity planes of three planets in Figure \ref{figpe_final}. Specifically, in Figure \ref{figpe_final}, top panels are for Model 3pA1 and bottom panels are for Model 3pA2. The left panels are for Kepler-396 b, and the middle and right panels are for Kepler-396 c and d respectively. The gray points are the values of the periods and eccentricities at the time $t=10^6$. In each panel, the black point with error bar presents the mean and standard deviation of the gray points, while the blue point with error bar shows the best-fit solution and its associated uncertainty, which is the standard deviation of the posterior distribution derived from the value listed in Table \ref{table:bestMCMC}. Among all six panels, the difference between the best-fit solution and the mean of 100 samples is not significant, comparing to the values of orbital periods or eccentricities. However, that difference 
is a bit larger for eccentricities of planet c in middle panels.
These results indicate that all 200 orbits remain dynamically stable throughout the simulations. 

Moreover, we calculate the Mean Exponential Growth factor of Nearby Orbits (MEGNO) indicators \citep{Cincotta2003} for these 200 simulations, and 
the result shows that all their MEGNO indicators are greater than 4 at $10^6$ years. Therefore, all these 200 orbits, which have initial conditions close to the best-fit solutions, are chaotic. As a result, while the Kepler-396 system  
should be dynamically stable during coming $10^6$ years, 
it could be moving on a chaotic orbit in the end.

On the other hand,
from the period ratio, it is known that the planet Kepler-396 b and the planet Kepler-396 c are close to 2:1 mean motion resonance. It is also found that the outer two planets, Kepler-396 c and Kepler-396 d are close to the 3:2 mean motion resonance. 
The normalized distance to $j:j-1$ resonance ($\Delta$) is defined as \cite{Lithwick2012}
\begin{equation}
\Delta=\frac{P_{\rm out}}{P_{\rm in}} \frac{j-1}{j}-1,\label{eq:delta}
\end{equation}
where $P_{\rm out}$ and  $P_{\rm in}$ are the periods of outer planet and inner planet respectively.  A planet pair, planet b and c, that has a period ratio of 2.06 is at a distance of $\Delta_{bc}=0.03$ from the 2:1 resonance, Meanwhile,
for planet pair, planet c and d, the normalized distance to
3:2 resonance is $\Delta_{cd}=0.08$. Therefore, there is a possibility that these three planets are in a 
resonant chain. Considering that Kepler-396 b, c, d are in a 1:2:3 resonant chain, the resonant angles, $\theta_i$ ($i = 1,\cdots, 4$), apsidal difference angles, and Laplace angle have the following form \citep{Kyriaki+2022AA}:
\begin{align}
&\theta_1 =2\lambda_c-\lambda_b-\varpi_b,\label{eq:a1}  \\
&\theta_2 =2\lambda_c-\lambda_b-\varpi_c,\label{eq:a2}  \\
&\theta_3 =3\lambda_d-2\lambda_c-\varpi_d,\label{eq:a3} \\ 
&\theta_4 =3\lambda_d-2\lambda_c-\varpi_c,\label{eq:a4} \\
&d{w_{cb}}=\varpi_c-\varpi_b,\label{eq:a5} \\
&d{w_{dc}}=\varpi_d-\varpi_c,\label{eq:a6} \\
&\phi_L =3\lambda_d-4\lambda_c+\lambda_b,\label{eq:a7}
\end{align}
where $\lambda_b$, $\lambda_c$, $\lambda_d$ are the  
mean longitudes of Kepler-396 b, c, d, respectively;
$\varpi_b$, $\varpi_c$, $\varpi_d$,
are the longitude of Kepler-396 b, c, d, respectively;
$d{w_{cb}}$ is the apsidal difference angle between 
Kepler-396 b and c, $d{w_{dc}}$ is the apsidal difference
angle between Kepler-396 c and d, and $\phi_L$ is the Laplace angle.
Whether any of the above angles are confined within certain ranges is key to confirming the resonant chain.   
For example, \cite{MacDonald2022K2-138} performed dynamical simulations and examined the libration of resonant angles. They found that a significant fraction of simulations led to libration, confirming K2-138 to be a 3:2 resonant chain.  
Moreover, \cite{Jontof2016} homogeneously performed TTV fitting on many resonant multi-planet systems. 
Both the mass-eccentricity degeneracy and apsidal alignment  were investigated. 

In order to examine whether our angles 
in Eq. (\ref{eq:a1})-(\ref{eq:a7}) are librating, we determine the values of resonant angles as functions of time with a reasonable time sampling rate during the $10^6$ years of all our 100 simulations.
Following the method in \cite{Wang2024AA}, the spreading level of angles can be estimated quantitatively. 
If the angle spreads all over 360 degrees, the probability that angle's values fall into one of the 
36 equal-sized bins between 0 and 360 degrees would be close
to the expected probability of a uniform spreading.
Thus, we use a chi-square values to measure the above mentioned closeness. 
After dividing 360 degrees into 36 equal-sized bins, 
the $\chi^2_{R}(\theta)$ of different resonant angle $\theta$ in 
any of our simulation is computed as
\begin{equation}
    \chi^2_{R}(\theta)=\Sigma^{36}_{i=1} \frac{(O_i(\theta)-E_i)^2}{E_i}, \label{eq:chi2theta}
\end{equation}
where $O_i(\theta)$ is how many times when the value of $\theta$ lies within the $i$-th bin in the simulation, and $E_i$ is the expected number under the assumption of a uniform spreading for the same simulation. 
Empirically, \cite{Wang2024AA} suggested that:
(a) when $\chi^2_{R}(\theta) > 1$, the resonant angle $\theta$ is in libration and the system is deeply inside the resonance; (b) when $\chi^2_{R}(\theta)< 0.1$, the resonant angle $\theta$ circulates, showing that the system is not in resonance; 
(c) when $ 0.1 <\chi^2_{R}(\theta)< 1$, the system is affected by the resonance, but the resonant angle $\theta$ may experience both the libration and circulation in a long time span.  From our simulations the values 
$\chi^2_{R}(dw_{cb})$ and $\chi^2_{R}(dw_{dc})$ are larger than one for all our 200 dynamical simulations, and the majority of other chi-square values of different resonant angle are less than 0.1. 

On the other hand, We also explicitly examined the evolution of all resonant angles. Starting from an initial condition with the parameters of Model 3pA1 
and Model 3pA2 in Table \ref{table:bestMCMC}, 
the values of resonant angles 
are determined as functions of time. Thus,
Figure \ref{fig:Ax_theta} in Appendix A presents the evolution of resonant angles, apsidal difference angles, and the Laplace angle of the best-fit solutions, 
i.e. Model 3pA1 and Model 3pA2. 
In order to view the angle libration, the values of all angles are plotted in the range [-180, 180] degrees here.   
It is clear that both $d{w_{cb}}$ and $d{w_{dc}}$ are confined to certain ranges, i.e. between -50 and 50 degrees approximately, so they are librating angles. 
In contrast, other resonance angles go from -180 to 180 degrees throughout. 

To further understand the libration ranges of the apsidal angles $d{w_{cb}}$ and $d{w_{dc}}$ of
all our 200 dynamical simulations, we determine the value
$RANGE_{cb}$ ($RANGE_{dc}$) which is the difference between the maximum and minimum values of $d{w_{cb}}$ ($d{w_{dc}}$) during entire $10^6$ years of simulations 
of Model 3pA1 and Model 3pA2. 
The histograms of $RANGE_{cb}$ and $RANGE_{dc}$
are presented in Figure \ref{fig:Range_dw}.
The libration ranges are between 100 and 113 degrees approximately. This confirms that the apsidal angles $d{w_{cb}}$ and $d{w_{dc}}$ of all 200 simulations are
librating angles as the ones explicitly shown
in Figure \ref{fig:Ax_theta}. 
Based on these results, it is validated that
three planets of the Kepler-396 system are in a three-body
apsidal resonance chain.

\section{Conclusions}
\label{sec:conclusions}

In this paper, we carefully study the architecture of the Kepler-396
planetary system and the properties of its planets, taking 
the advantage of new TESS data.
After the data analysis which employing the PLD technique, the light curves are fitted and mid-transit times are determined. 
With a long baseline TTV, the parameter space of the mass and orbit
of the third planet is explored. The best-fit three-planet models with 
the reduced-chi-square values $\chi_{red}^2=2.24$ and $\chi_{red}^2=2.34$
are found, which are significantly smaller than the value $\chi_{red}^2=12.97$ of two-planet model. Therefore, the existence of the third planet Kepler-396 d is 
strongly suggested.

The above two best-fit three-planet models with very similar reduced-chi-square values are obtained
through different prior of orbital inclinations. 
Their values of best-fit parameters are
very close except the inclinations. Thus, it is clear that 
the current observational data can only provide these best-fit results with the unsettled inclination. 

Based on these two best-fit solutions, 
the masses of Kepler-396 b, c, d are updated to be around
9, 4, 4 $M_\earth$,
and the corresponding orbital periods are
43, 89, 144 days, respectively.
Thus, Kepler-396 b is on the border
between rocky or volatile-rich, 
Kepler-396 c is confirmed to be a volatile-rich super-puff planet, and both planets lie in the Neptune savanna. 

Finally, two hundred dynamical simulations derived from the best-fit solutions with three planets are performed for $10^6$ years. These three planets
are stable during the simulations. The resonant angles among three planets 
during the above simulations are studied. Our results show that they are in a three-body apsidal resonance chain.
This newly confirmed apsidal resonance chain with relatively wider orbits could become an interesting laboratory 
for further investigation on the formation of resonant exoplanet systems. We conclude that the Kepler-396 system 
needs further studies with future high quality observational 
data in order to understand the orbital inclinations and 
the formation of this three-body apsidal resonance chain.

\vspace{0.5cm}
\section*{Acknowledgments}

We are indebted to the anonymous referee for 
good comments and suggestions, which improve this paper significantly.
This work is supported by the grant from the National Science and Technology Council (NSTC), Taiwan. The grant numbers are NSTC 113-2112-M-007-030, NSTC 114-2112-M-007-029, and 
115-2112-M-007-009. 
TESS data presented in this paper were obtained from the Mikulski Archive for Space Telescopes (MAST) at the Space Telescope Science Institute
(STScI). The specific
observations analyzed can be accessed via 
\dataset[https://doi.org/10.17909/t9-nmc8-f686]{https://doi.org/10.17909/t9-nmc8-f686}. 


\vspace{5mm}
\facilities{Transiting Exoplanet Survey Satellite (TESS)}
\software{{\tt\string nuance} \citep{Garcia+2024AJ},
{\tt\string Lightkurve}\citep{Lightkurve2018},
{\tt\string juliet} \citep{Espinoza2019JULIET},
{\tt \string dynesty} \citep{Speagle2020MNRAS}, 
{\tt\string TTVFast} \citep{Deck2014}, 
{\tt\string SciPy} \citep{Virtanen2020}, 
{\tt\string emcee} \citep{Foreman2013}, 
{\tt\string PyLightcurve} \citep{Tsiaras2016},
{\tt\string REBOUND} \citep{ReinLiu2012}.
}

\bibliography{K396}{}
\bibliographystyle{aasjournal}

\appendix

\FloatBarrier
\counterwithin{figure}{section}
\section{Figures}

\begin{figure*}
\centering
\hspace{-2.9cm}
\includegraphics[width=0.9\textwidth]{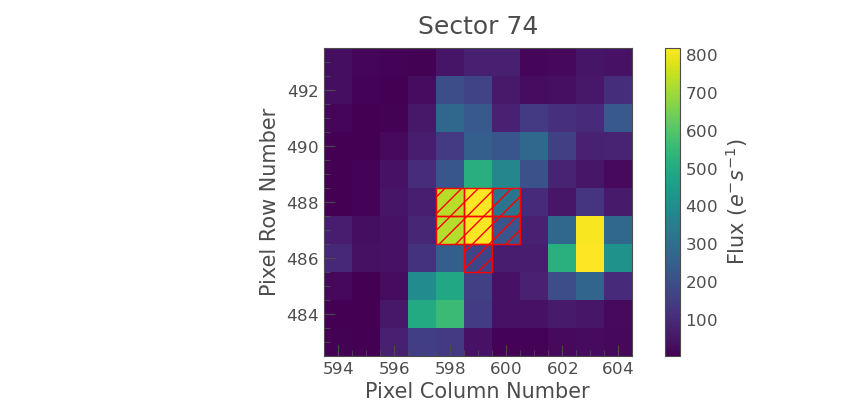}
\caption{The TPF of Kepler-396 at the starting time of Sector 74. The red slashed region is chosen as the aperture.}
\label{fig_tpf}
\end{figure*}

\begin{figure*}
\centering
\includegraphics[width=1.0\textwidth]{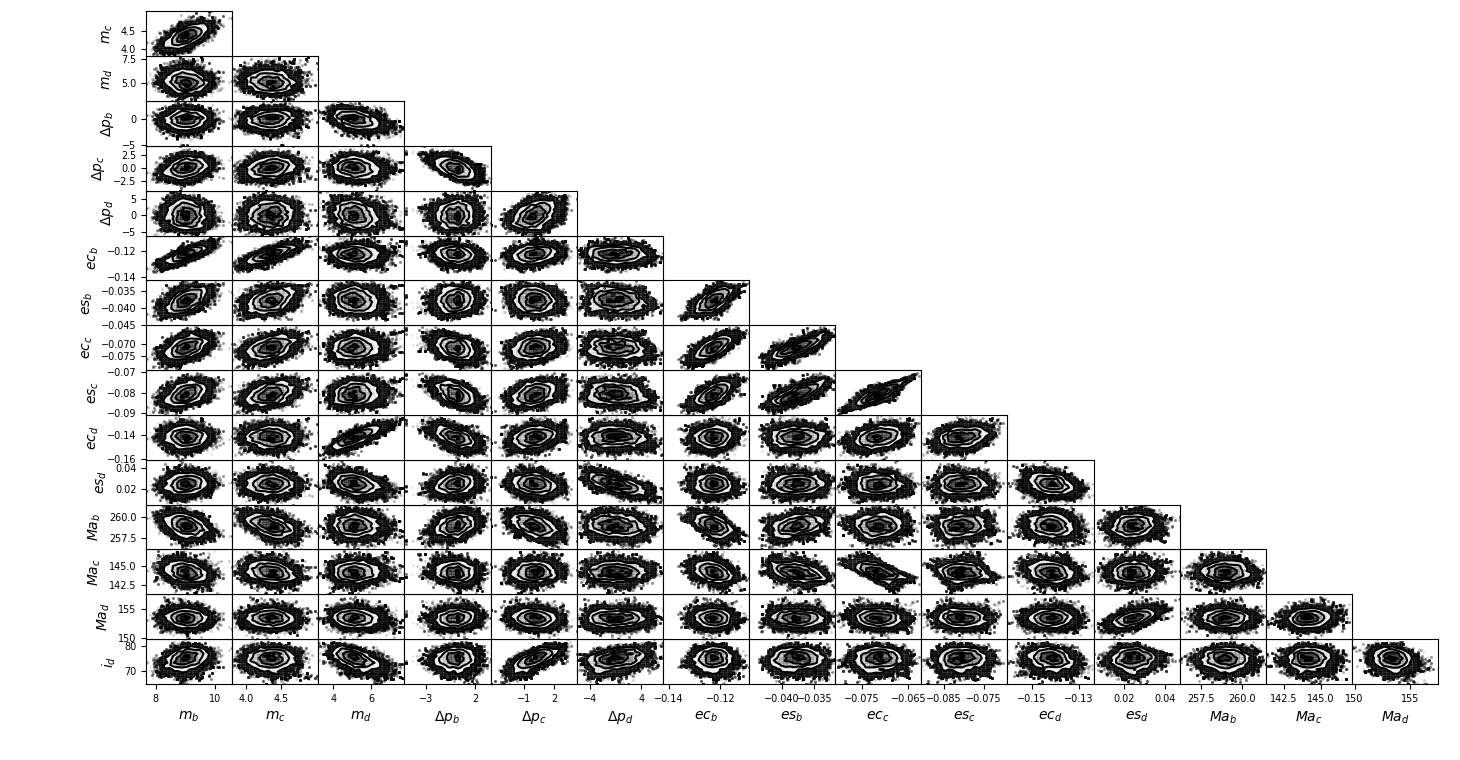}
\caption{The posterior distributions of the parameters of MCMC samples, i.e. corner plot, for Model 3pA1, where $\Delta p_b=(p_b-42.99415)\times 10^4$, $\Delta p_c=(p_c-88.50859)\times 10^3$, and $\Delta p_d=(p_d-144.0025)\times 10^2$  }
\label{fig:corner_Ax}
\end{figure*}

\begin{figure*}
\centering
\includegraphics[width=0.80\textwidth]{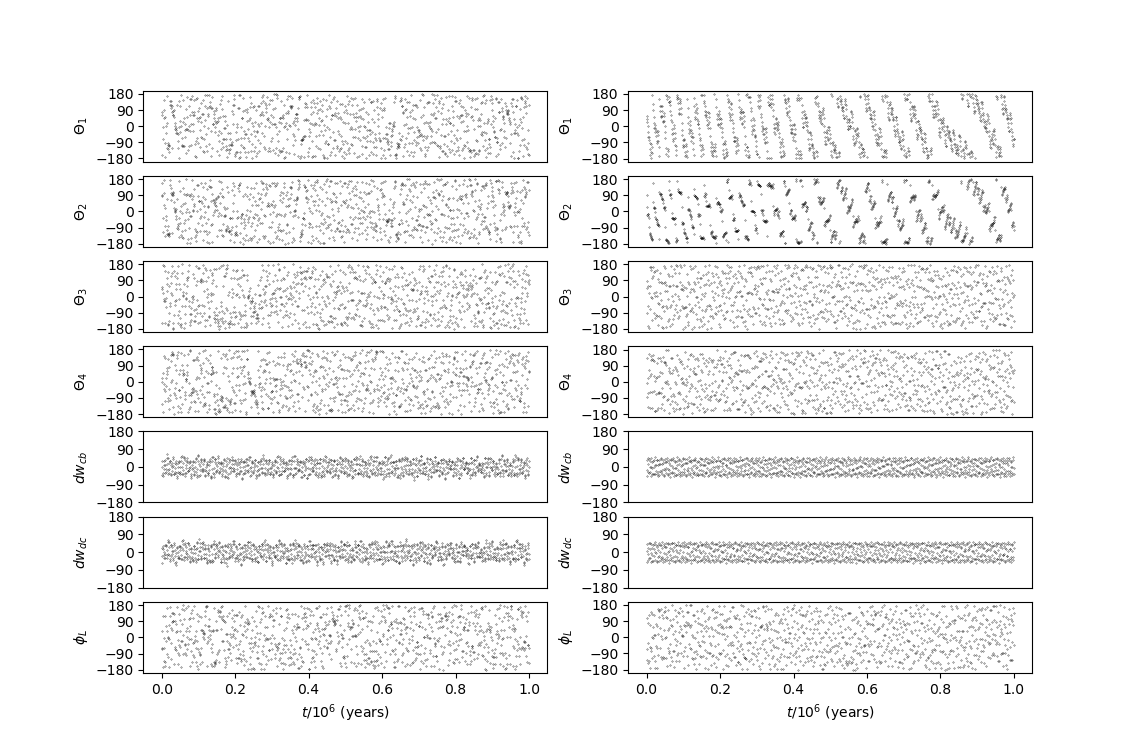}
\caption{The resonant angles, apsidal difference angles and Laplace angles of the best-fit solutions in Model 3pA1 (left panel) and Model 3pA2 (right panel) over $10^6$ years.}
\label{fig:Ax_theta}
\end{figure*}

\begin{figure*}
\centering
\includegraphics[height=0.45\textheight]{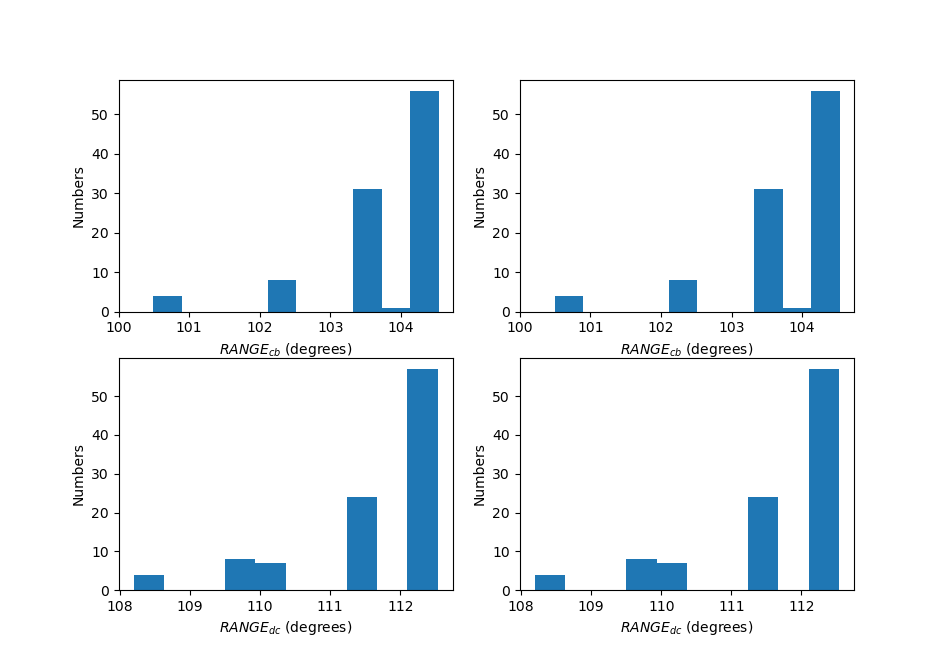}
\caption{The histograms of $RANGE_{cb}$ and $RANGE_{dc}$   for the 100 simulations of Model 3pA1 (left panels) and
another 100 simulations of Model 3pA2 (right panels).}
\label{fig:Range_dw}
\end{figure*}

\FloatBarrier
\counterwithin{table}{section}
\section{The Tables of Mid-Transit Times}

\begin{table*}
\caption{The mid-transit times of the planet Kepler-396 b. The epoch IDs, mid-transit times, data sources, and references are listed in different columns successively.}
\label{tab:Planet b}
\hspace{0.9cm}
\begin{tabular}{|c|c|c|c|}\hline
Epoch ID&  T (BJD$_{\rm TDB}$) & Data Source & Reference  \\ \hline
0 & $2454995.5046550 \pm  0.0027128$  &Kepler & \cite{Thompson2018}\\
1 & $2455038.5056457 \pm  0.0043931$  &Kepler & \cite{Thompson2018}\\
2 & $2455081.4929876 \pm  0.0018318$ &Kepler & \cite{Thompson2018}\\
3 & $2455124.4832444 \pm  0.0038654$   &Kepler & \cite{Thompson2018}\\
4 & $2455167.4786777 \pm  0.0012499$    &Kepler & \cite{Thompson2018}\\
5 & $2455210.4751203 \pm  0.0038194$   &Kepler & \cite{Thompson2018}\\
6 & $2455253.4561320 \pm  0.0040595$   &Kepler & \cite{Thompson2018}\\
7 & $2455296.4588458 \pm  0.0034318$   &Kepler & \cite{Thompson2018}\\
8 & $2455339.4395981 \pm  0.0026923$   &Kepler & \cite{Thompson2018}\\
9 & $2455382.4349892 \pm  0.0022908$   &Kepler & \cite{Thompson2018}\\
10 &$2455425.4203794 \pm  0.0022872$    &Kepler & \cite{Thompson2018}\\
11 &$2455468.4009739 \pm  0.0025360$   &Kepler & \cite{Thompson2018}\\
12 &$2455511.3960510 \pm  0.0036016$    &Kepler & \cite{Thompson2018}\\
14 &$2455597.3769857 \pm  0.0023602$   &Kepler & \cite{Thompson2018}\\
16 &$2455683.3609161 \pm  0.0042476$    &Kepler & \cite{Thompson2018}\\
17 &$2455726.3546024 \pm  0.0018195$   &Kepler & \cite{Thompson2018}\\
18 &$2455769.3450915 \pm   0.0016349$    &Kepler & \cite{Thompson2018}\\
19 &$2455812.3485682 \pm   0.0066320$    &Kepler & \cite{Thompson2018}\\
20 &$2455855.3360638 \pm 0.0042448$   &Kepler & \cite{Thompson2018}\\
21 &$2455898.3312578 \pm  0.0029204$    &Kepler & \cite{Thompson2018}\\
22 &$2455941.3232484 \pm   0.0038981$    &Kepler & \cite{Thompson2018}\\
23 &$2455984.3157597 \pm  0.0017386$   &Kepler & \cite{Thompson2018}\\
24 &$2456027.3114447 \pm   0.0026943$    &Kepler & \cite{Thompson2018}\\
25 &$2456070.3162920 \pm   0.0022518$    &Kepler & \cite{Thompson2018}\\
26 &$2456113.3051406 \pm  0.0019665$   &Kepler & \cite{Thompson2018}\\
27 &$2456156.2980625 \pm   0.0051616$   &Kepler & \cite{Thompson2018}\\
28 &$2456199.3034817 \pm   0.0032386 $   &Kepler & \cite{Thompson2018}\\
29 &$2456242.2963682 \pm   0.0040752$  &Kepler & \cite{Thompson2018}\\
30 &$2456285.2984696 \pm   0.0034537$   &Kepler & \cite{Thompson2018}\\
31 &$2456328.2923114 \pm  0.0021861 $  &Kepler & \cite{Thompson2018}\\
32 &$2456371.2930071 \pm  0.0022548 $  &Kepler & \cite{Thompson2018}\\
33 &$2456414.2972786 \pm  0.0050519 $ &Kepler & \cite{Thompson2018}\\
86 &$2458692.9172^{+0.0069}_{-0.0060}$ & TESS & \cite{Jontof2022}\\ 
87 &$2458735.8870^{+0.047}_{-0.036}$  & TESS & \cite{Jontof2022} \\
103&$2459423.8437^{+0.0077}_{-0.0070}$ & TESS & \cite{Jontof2022}\\
124& $2460326.6200\pm {0.0582}$ & TESS & this work \\ 
125&$2460369.6727\pm 0.0288 $  & TESS & this work\\\hline
\end{tabular}
\end{table*}

\begin{table*}
\caption{The mid-transit times of the planet Kepler-396 c. The epoch IDs, mid-transit times, data sources, and references are listed in different columns successively.}
\label{tab:Planet c}
\hspace{0.85cm}
\begin{tabular}{|c|c|c|c|}\hline
Epoch ID&  T (BJD$_{\rm TDB}$) & Data Source & Reference  \\ \hline
1&  $2455104.1775816  \pm 0.0011803$ & Kepler & \cite{Thompson2018} \\
2 & $2455192.6831741  \pm 0.0029773$ &Kepler & \cite{Thompson2018}   \\
3 & $2455281.1953293 \pm  0.0039428$  &Kepler &  \cite{Thompson2018}  \\
4 & $2455369.7129499 \pm  0.0035767$  &Kepler & \cite{Thompson2018}   \\
5 & $2455458.2365655 \pm  0.0026320$  &Kepler &  \cite{Thompson2018}   \\
6 & $2455546.7562663 \pm  0.0018285$ & Kepler &\cite{Thompson2018} \\      
7 & $2455635.2698463 \pm  0.0030252$ &Kepler &\cite{Thompson2018} \\       
8 & $2455723.8038602 \pm  0.0016501$ &Kepler &  \cite{Thompson2018}\\
9 & $2455812.3319296 \pm  0.0038559$ &Kepler & \cite{Thompson2018}\\       
10 &$2455900.8309857 \pm  0.0013251$ &Kepler & \cite{Thompson2018}\\       
11 &$2455989.3354928 \pm  0.0038198$&Kepler & \cite{Thompson2018}\\      
12 &$2456077.8139020 \pm  0.0046143$ &Kepler & \cite{Thompson2018} \\
13 &$2456166.3118562 \pm  0.0013619$  &Kepler & \cite{Thompson2018}\\
14 &$2456254.8051466 \pm  0.0012024$ &Kepler & \cite{Thompson2018}\\       
15 &$2456343.2963388 \pm  0.0040067$ &Kepler & \cite{Thompson2018}\\      
42 &$2458733.3103^{+0.0064}_{-0.0039}$   & TESS & \cite{Jontof2022}\\
50 &$2459441.2730^{+0.0079}_{-0.0061}$   & TESS & \cite{Jontof2022}\\
54 &$2459795.3749\pm 0.0221$  & TESS & this work \\
60 &$2460326.5849\pm 0.0286$ & TESS & this work\\
\hline
\end{tabular}
\end{table*}

\end{document}